# An Isotropic Discretization with Semi-implicit Approach for Phase Field Model of Alloy Solidification

C. Tang*, D.T. Wu, S.S. Quek

Institute of High Performance Computing (IHPC), Agency for Science Technology and Research (A*STAR), 1 Fusionopolis Way, #16-16 Connexis, Singapore 138632, Republic of Singapore

**Abstract**

Quantitative phase field models have been extensively used to study the solidification behavior of alloys under different conditions. However, a longstanding challenge of phase field models is the directional bias caused by the discretization-induced lattice effects. In particular, widely used discretization methods may introduce significant spurious anisotropy for simulations of polycrystalline solidification. In this paper, we demonstrate a feasible 2D discretization strategy utilizing a hexagonal mesh to reduce the lattice-induced anisotropy of the phase field model. The leading differential terms of the 2D discretization methods are analyzed by using known methods in Fourier space. Using Taylor expansion of discrete Fourier Transform up to sixth order, we found that the proposed discretization strategy is more accurate and isotropic than other methods, including the isotropic discretization recently proposed by Ji *et al.*[1]. Additionally, the proposed 2D discretization method can be easily incorporated into a semi-implicit algorithm to solve phase field equations, thereby greatly reducing time step constraints and improving computational efficiency compared to explicit approaches. To prove the accuracy and efficiency of the proposed isotropic discretization with semi-implicit algorithm, 2D simulations of alloy solidification with different discretization schemes were performed and compared. We show that the proposed discretization using a hexagonal mesh can drastically reduce grid-induced anisotropy compared to conventional methods.

**1. Introduction**

Over the past three decades, phase field (PF) models have emerged as popular approaches to predicting the microstructure formation of pure metal and multicomponent alloys [2-6]. The basic concept of PF models is to employ a scalar variable $\phi$ that smoothly varies across the solid/liquid interface, thus avoiding the explicit tracking of the physical interface with complex configuration [7]. Accordingly, a set of non-linear partial differential equations (PDEs) are numerically solved to track the temporal evolution of interface and other quantities (e.g., temperature and solute concentration) [8]. For alloy solidification, Karma and coworkers [9, 10] proposed a PF model to quantitatively simulate the dendritic solidification of dilute binary systems with arbitrary interfacial thickness. By introducing the so-called antitrapping term in the solute transport equation, mesoscale interfacial thickness can be chosen to efficiently simulate dendrite formation under isothermal or non-isothermal conditions. Recently, this well-established PF model has proven to be useful in many applications, such as isothermal crystallization [11], directional solidification [12-14], and competitive solidification during additive manufacturing [15-17].

Finite difference discretizations using 2D square or 3D cubic lattices have been conventionally applied to solve the PDEs of PF models. In most simulation studies, standard finite difference discretization is applied to compute the associated spatial derivatives that appeared in the PF model, except for Laplacian terms [18] and second-order spatial derivatives in the solute transport equation. Instead of using the simplest form, many researchers discussed the application of isotropic discretization to compute the Laplacian. For example, 5-point stencil discretization is the simplest approach to compute Laplacian on a 2D square lattice, yet it is known as an anisotropic scheme [19]. In comparison, 9-point stencil discretization (Eqn. (B1) in Ref. [10]) was shown to converge towards rotational invariance with $h^2$ order. Regarding the divergence of solute flux in the solute transport equation, finite volume discretization with a 5-point stencil (Appendix B of Ref. [10]) is commonly used for 2D simulation. For the divergence of antitrapping current, a special 9-point stencil discretization is typically adopted (Appendix B of Ref. [10]). However, these widely adopted discretization methods can still introduce spurious anisotropy in polycrystalline solidification, especially for typical simulations where grid spacing ($h$) is comparable to the chosen interfacial thickness ($W$). In a recent work, Ji *et al.* [1] proposed isotropic discretizations on 2D square and 3D cubic lattices. Their work suggests that the discretization of solute diffusion and compensating antitrapping terms is crucial to circumvent lattice-induced anisotropy. Their elegant way to discretize solute diffusion and the antitrapping term is attractive for the accurate prediction of polycrystalline alloy solidification. To the best of our knowledge, there are few studies successfully addressing the lattice effects due to the discretization of solute diffusion and antitrapping current. Unfortunately, the implementation of the isotropic discretization proposed by Ji *et al.* [1] is not straightforward for the PF equations. Considering the complexity of their isotropic discretization, it would be challenging to develop an implicit algorithm to solve the corresponding PDEs. Consequently, the adoption of explicit approach can impose a time-step constraint on the numerical algorithms of the PF model. Such a constraint would be undesirable for efficient simulations of polycrystalline solidification.

In this work, we present a feasible 2D discretization strategy based on a hexagonal mesh to compute the second-order spatial derivatives in the PF equations. The discretization method is a standard form of finite volume discretization and thus can be easily incorporated into a semi-implicit numerical algorithm. The rotational invariance of this discretization was examined in Fourier space and compared with a previous study by Ji *et al.* [1]. To verify the accuracy and efficiency of the proposed discretization, we also performed 2D simulations of dendritic solidification under isothermal and non-isothermal conditions with different schemes. Simulation results suggest that the proposed isotropic discretization with semi-implicit algorithm can greatly reduce lattice effects in 2D polycrystalline solidification simulations. The accuracy of our proposed discretization strategy is better than the conventionally used method and the isotropic strategy from Ji *et al.* [1]. Additionally, compared to the explicit algorithm in Ref. [1], the time step constraint is reduced by one to two orders of magnitude using the semi-implicit algorithm that is readily implementable by our proposed discretization method.

## 2. The phase field model

We employed a well-known PF model [9, 10] to simulate the dendritic solidification of a dilute binary alloy. In 2D, the governing equations of the PF model are given as

$$\tau_0 a_s^2 [1+(1-k_e)U]\frac{\partial \phi}{\partial t} = W^2 \nabla \cdot (a_s^2 \nabla \phi) + W^2 \partial_x (|\nabla \phi|^2 a_s \frac{\partial a_s}{\partial \phi_x}) + W^2 \partial_y (|\nabla \phi|^2 a_s \frac{\partial a_s}{\partial \phi_y})$$
$$+ (\phi - \phi^3) - \lambda (1-\phi^2)^2 \left[ U + \frac{T-T_s}{|m_l|c_l^0(1-k_e)} \right] \quad (1)$$

$$\frac{1}{2}[1+k_e-(1-k_e)\phi]\frac{\partial U}{\partial t}=\nabla\cdot(\frac{1-\phi}{2}D_l\nabla U)+\frac{1}{2}[1+(1-k_e)U]\frac{\partial \phi}{\partial t}$$
$$+\nabla\cdot\left\{[1+(1-k_e)U]\frac{W}{2\sqrt{2}}\frac{\partial \phi}{\partial t}\frac{\nabla\phi}{|\nabla\phi|}\right\} \quad (2)$$

where the order parameter $\phi$ (ranges from -1 to +1) is used to examine the temporal evolution of solid/liquid interface, $U$ represents the dimensionless supersaturation and is associated with the solute concentration $c$, $t$ is time, $T$ is temperature, $T_l$ is the liquidus temperature with an equilibrium solute concentration of liquid $c_l^0$, $\tau_0$ is the characteristic relaxation time, $m_l$ is the liquidus slope, $D_l$ is the chemical diffusivity of solute atoms in liquid, $k_e$ is solute partition coefficient, $W$ is the arbitrarily chosen interfacial thickness, $\lambda$ is a coupling parameter depending on $W$, $a_s$ is the strength of anisotropy.

For cubic alloy systems with fourfold symmetry in 2D, the strength of anisotropy is described by

$$a_s=1-3\varepsilon_4+4\varepsilon_4\cdot\frac{\phi_x^4+\phi_y^4}{(\phi_x^2+\phi_y^2)^2} \quad (3)$$

where $\varepsilon_4$ denotes the anisotropy of interfacial energy.

The dimensionless supersaturation $U$ is given as

$$U=\left[\frac{2c}{1+k_e-(1-k_e)\phi}-c_l^0\right]\Big/\left(c_l^0-k_ec_l^0\right) \quad (4)$$

As suggested by Ji *et al.* [1], isotropic discretization of the second-order spatial derivatives in Eqns. (1) and (2) are important to eliminate lattice effects in PF simulations.

For the evolution of $\phi$, the leading differential term in Eqn. (1) should be

$$T_1(\vec{r})=\nabla\cdot(a_s^2\nabla\phi) \quad (5)$$

Note that this leading term is different from the study by Ji *et al.* [1], since they used the preconditioned field variable, $\psi$, instead of the order parameter, $\phi$, used in this work.

For the solute transport equation, the leading differential terms of Eqn. (2) include the divergence of solute flux

$$T_2(\vec{r}) = \nabla \cdot (\frac{1-\phi}{2} D_l \nabla U) \tag{6}$$

and the divergence of the antitrapping current,

$$T_3(\vec{r}) = \nabla \cdot \left\{ [1+(1-k_e)U] \frac{W}{2\sqrt{2}} \frac{\partial \phi}{\partial t} \frac{\nabla \phi}{|\nabla \phi|} \right\} \tag{7}$$

Consequently, the leading differential terms of PF equations can be taken as generalized forms $\nabla \cdot (\alpha \nabla \beta)$ and $\nabla \cdot (\alpha \frac{\nabla \beta}{|\nabla \beta|})$, where $\alpha$ and $\beta$ represent arbitrary scalar variables. Although Eqn. (7) with the original antitrapping term can be computed with an isotropic discretization, a complicated 21-point stencil (Appendix B of Ref. [1]) is required to get the isotropic form of $\nabla \cdot (\alpha \frac{\nabla \beta}{|\nabla \beta|})$ on a 2D square lattice. Therefore, Ji *et al.* [1] presented an approximated antitrapping term with the generalized form of $\nabla \cdot (\alpha \nabla \beta)$. As a result, fewer stencil points are required to compute the isotropic discretization of the approximated antitrapping term. Guided by their approximation of the antitrapping current, we propose a different approximation of the antitrapping term with the scalar variable $\phi$.

In Karma's model, the equilibrium profile of $\phi$ is a hyperbolic tangent [10]

$$\phi_0(r) = -\tanh(\frac{r}{\sqrt{2}W}) \tag{8}$$

Therefore, the gradient magnitude of $\phi_0$ is

$$|\nabla \phi_0| = \left|\frac{\partial \phi_0}{\partial r}\right| = \frac{|1-\phi_0^2|}{\sqrt{2}W} \tag{9}$$

By replacing $|\nabla \phi|$ in Eqn. (7) with $|\nabla \phi_0|$, the approximated antitrapping term in this work is given as

$$\overline{T}_3(\vec{r}) = \nabla \cdot \left\{ \frac{[1+(1-k_e)U]W^2}{2} \frac{1}{|1-\phi^2|+\varepsilon_{small}} \frac{\partial \phi}{\partial t} \nabla \phi \right\} \tag{10}$$

where $\varepsilon_{small}$ (0.001 in this study) is a small number to avoid division by zero. Our approximated antitrapping term also has a generalized form of $\nabla \cdot (\alpha \nabla \beta)$, for which it is easier to achieve isotropic discretization using fewer stencil points compared to Eqn. (7). Our numerical algorithm deals with PF equations with Eqn. (10) as the approximated antitrapping current.

## 3. Isotropic discretization of the leading differential terms

In this paper, we used the notations $L$ and $\tilde{D}$ to represent respectively, the discretization of Laplacian ($\nabla^2 \beta$) and generalized divergence ($\nabla \cdot (\alpha \nabla \beta)$). As discussed above, all the leading differential terms in the PF model can be expressed as the generalized form $\nabla \cdot (\alpha \nabla \beta)$. Therefore, the discretization of generalized divergence $\nabla \cdot (\alpha \nabla \beta)$ is mainly discussed in this section. Compared to the isotropic discretization $\tilde{D}_{2,1}$ proposed by Ji *et al.* [1], our proposed discretization can provide better isotropy for 2D PF simulations. Details of Ji's $\tilde{D}_{2,1}$ discretization [1] is given in Appendix A. Since many researchers require the discretization of Laplacian in their PF simulations (e.g., Steinbach's PF model [8]), we briefly discuss the discretization of Laplacian ($\nabla^2 \beta$) and show its rotational invariance.

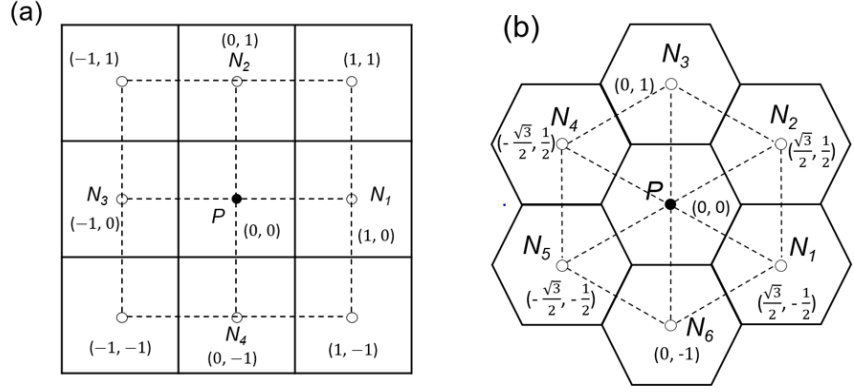

Figure 1. Schematic of 2D finite volume cells with (a) square and (b) hexagonal shapes. The dashed lines represent the equivalent finite difference lattices.

Ji *et al.* [1] stated that $\tilde{D}_{1,0}$ discretization using 9-point stencil (see Appendix A) is commonly used to compute $\nabla \cdot (\alpha \nabla \beta)$ in 2D square grids. However, it is worth mentioning that their claim regarding the $\tilde{D}_{1,0}$ discretization may be misleading because the most widely used method for $\nabla \cdot (\alpha \nabla \beta)$ computation is the standard finite volume discretization with a 5-point stencil [10, 12]. For an arbitrary mesh, the standard finite volume discretization of $\nabla \cdot (\alpha \nabla \beta)$ may be found elsewhere (Eqn. (3.24) in Ref. [20]). For 2D square grids, the grid spacing $h_{sq}$ is the distance between the centroids of two nearest neighbours. As illustrated in Fig. 1(a), the standard finite volume discretization of $\nabla \cdot (\alpha \nabla \beta)$ at the centroid of grid cell $P$ is expressed as

$$\nabla \cdot (\alpha \nabla \beta)\big|_P = \frac{1}{A_{sq}} \sum_{N=1}^{4} L_{sq} \alpha_{f,N} \frac{(\beta_N - \beta_P)}{h_{sq}} \qquad (11)$$

where $L_{sq} = h_{sq}$ is the edge length of the square grid, $A_{sq} = h_{sq}^2$ is the area of the square grid, points $N$ represents the centroids of four nearest neighbours of cell $P$, and $\alpha_{f,N}$ represents the $\alpha$ values at the mid-point of the edge of mesh grid *P*. Linear interpolation is typically used to calculate $\alpha_{f,N}$. For square grids, $\alpha_{f,N}$ is thus given by

$$\alpha_{f,N} = \frac{\alpha_P + \alpha_N}{2} \tag{12}$$

Following a similar way in Ref. [1] to express $\nabla \cdot (\alpha \nabla \beta)$ at point (0, 0), the standard finite volume discretization of generalized divergence in a 2D square mesh is

$$\begin{aligned}
\nabla \cdot (\alpha \nabla \beta) &= \frac{1}{2h_{sq}^2}\left[(\alpha_{(1,0)} + \alpha_{(0,0)})(\beta_{(1,0)} - \beta_{(0,0)})\right] \\
&+ \frac{1}{2h_{sq}^2}\left[(\alpha_{(0,1)} + \alpha_{(0,0)})(\beta_{(0,1)} - \beta_{(0,0)})\right] \\
&+ \frac{1}{2h_{sq}^2}\left[(\alpha_{(-1,0)} + \alpha_{(0,0)})(\beta_{(-1,0)} - \beta_{(0,0)})\right] \\
&+ \frac{1}{2h_{sq}^2}\left[(\alpha_{(0,-1)} + \alpha_{(0,0)})(\beta_{(0,-1)} - \beta_{(0,0)})\right]
\end{aligned} \tag{13}$$

Eqn. (13) is the simplest approach to calculate generalized divergence $\nabla \cdot (\alpha \nabla \beta)$ in 2D square grids and can be readily implemented for an implicit or semi-implicit algorithm. Such a discretization is commonly used to calculate solute diffusion in the PF model [10, 12]. In the paper, this discretization is denoted as $\tilde{D}_{FV,sq}$, which is different from the $\tilde{D}_{1,0}$ discretization in Ji's work (Appendix A). As discussed later, the simple $\tilde{D}_{FV,sq}$ discretization is an anisotropic scheme for PF simulations. On the other hand, since $\tilde{D}_{1,0}$ in [1] is neither an isotropic form nor a common discretization strategy, further discussion of $\tilde{D}_{1,0}$ is omitted from this work.

In this work, the proposed isotropic discretization for $\nabla \cdot (\alpha \nabla \beta)$ computation is a standard form of finite volume discretization. Consequently, it is straightforward to implement the proposed schemes in an implicit manner. Instead of using squares, regular hexagons are used to generate finite volume cells in 2D space, as shown in Fig. 1(b). Note that the hexagonal finite volume mesh is equivalent to a finite difference lattice with regular triangles. As illustrated by the dashed lines in Fig. 1(b), the centroids of hexagonal cells are corresponding nodal points of triangular lattice for finite difference methods. Accordingly, our isotropic discretization method can be implemented with hexagonal finite volume cells or triangular finite difference lattices.

For hexagonal mesh, the grid spacing $h_{hex}$ is the distance between the centroids of two nearest hexagons. The standard finite volume discretization of generalized divergence $\nabla \cdot (\alpha \nabla \beta)$ is thus given as

$$\nabla \cdot (\alpha \nabla \beta)\big|_P = \frac{1}{A_{hex}} \sum_{N=1}^{6} L_{hex} \frac{(\alpha_N + \alpha_P)}{2} \frac{(\beta_N - \beta_P)}{h_{hex}} = \frac{1}{3h_{hex}^2} \sum_{N=1}^{6} (\alpha_N + \alpha_P)(\beta_N - \beta_P) \qquad (14)$$

where $L_{hex} = \frac{\sqrt{3}}{3} h_{hex}$ is the edge length of the hexagon, $A_{hex} = \frac{\sqrt{3}}{2} h_{hex}^2$ is the area of the hexagon. For instance, the discretization of generalized divergence $\nabla \cdot (\alpha \nabla \beta)$ in Fig. 1(b) is

$$\nabla \cdot (\alpha \nabla \beta) = \frac{1}{3h_{hex}^2}\left[(\alpha_{(\frac{\sqrt{3}}{2},\frac{1}{2})} + \alpha_{(0,0)})(\beta_{(\frac{\sqrt{3}}{2},\frac{1}{2})} - \beta_{(0,0)})\right]$$

$$+ \frac{1}{3h_{hex}^2}\left[(\alpha_{(0,1)} + \alpha_{(0,0)})(\beta_{(0,1)} - \beta_{(0,0)})\right]$$

$$+ \frac{1}{3h_{hex}^2}\left[(\alpha_{(-\frac{\sqrt{3}}{2},\frac{1}{2})} + \alpha_{(0,0)})(\beta_{(-\frac{\sqrt{3}}{2},\frac{1}{2})} - \beta_{(0,0)})\right]$$

$$+ \frac{1}{3h_{hex}^2}\left[(\alpha_{(-\frac{\sqrt{3}}{2},-\frac{1}{2})} + \alpha_{(0,0)})(\beta_{(-\frac{\sqrt{3}}{2},-\frac{1}{2})} - \beta_{(0,0)})\right] \quad (15)$$

$$+ \frac{1}{3h_{hex}^2}\left[(\alpha_{(0,-1)} + \alpha_{(0,0)})(\beta_{(0,-1)} - \beta_{(0,0)})\right]$$

$$+ \frac{1}{3h_{hex}^2}\left[(\alpha_{(\frac{\sqrt{3}}{2},-\frac{1}{2})} + \alpha_{(0,0)})(\beta_{(\frac{\sqrt{3}}{2},-\frac{1}{2})} - \beta_{(0,0)})\right]$$

Accordingly, the proposed isotropic discretization of generalized divergence (Eqn. (14) or (15)) is denoted as $\tilde{D}_{FV,hex}$.

For the discretization of Laplacian ($\nabla^2 \beta$) in square grids, the standard finite volume discretization (denoted as $L_{FV,sq}$) is identical to the $L_{1,0}$ discretization in Ji's study [1]. The standard finite volume discretization in regular hexagons (denoted as $L_{FV,hex}$) is an isotropic scheme, and its form is given as

$$\nabla^2 \beta \big|_P = \frac{2}{3h_{hex}^2}\sum_{N=1}^{6}(\beta_N - \beta_P) \quad (16)$$

For example, the $L_{FV,hex}$ discretization of $\nabla^2 \beta$ in Fig. 1(b) is

$$\nabla^2 \beta = \frac{2}{3h_{hex}^2}\left[\beta_{(\frac{\sqrt{3}}{2},\frac{1}{2})} + \beta_{(0,1)} + \beta_{(-\frac{\sqrt{3}}{2},\frac{1}{2})} + \beta_{(-\frac{\sqrt{3}}{2},-\frac{1}{2})} + \beta_{(0,-1)} + \beta_{(\frac{\sqrt{3}}{2},-\frac{1}{2})} - 6\beta_{(0,0)}\right] \quad (17)$$

Eqns. (14) and (16) are standard forms of finite volume discretization in 2D, which is typical for computational fluid dynamics (CFD) simulations. Accordingly, these two discretization methods can be readily implemented with a semi-implicit algorithm. Moreover, it is straightforward to implement the same discretization in finite difference lattice with regular triangles. Analyses and results of the proposed isotropic discretizations are discussed in the following section.

## 4. Fourier analysis and simulation results

4.1. Fourier analysis of isotropic schemes

The discrete Fourier Transform (DFT) can be applied to examine the leading anisotropic errors of various discretization methods [1, 21]. As discussed in previous studies, the anisotropy of discretization can be directly visualized in Fourier space for qualitative understanding. At a wave vector $\vec{k} = (k_x, k_y)$, the DFT of Laplacian has the form

$$L(\vec{k}) = \frac{\sum_{\vec{r}} e^{-i\vec{k}\cdot\vec{r}} L(\vec{r})}{\sum_{\vec{r}} e^{-i\vec{k}\cdot\vec{r}} \phi(\vec{r})} \tag{18}$$

The Fourier transform of Laplacian in continuous limit gives an isotropic form with $L(\vec{k}) = -k^2$ where $k = \sqrt{k_x^2 + k_y^2}$ [1]. The 5-point stencil discretization of Laplacian ($L_{FV,sq}(\vec{k}) = L_{1,0}(\vec{k})$) is considered anisotropic since the leading anisotropic error occurs at the fourth order of $k$ in the Taylor series of $L$ about $k = 0$. The $L_{2,1}$ scheme by Ji *et al.* [1] (given in Appendix A) exhibits better isotropy since rotational variance first appears at the sixth order of $k$ in the Taylor expansion of $L$ [1]. Details of the DFT of $L_{1,0}$ and $L_{2,1}$ schemes can be found in Eqns. (33) to (35) of Ref. [1]. For the standard finite volume discretization with hexagonal mesh, the DFT of our proposed method (Eqn. (17)) yields

$$L_{FV,hex}(\vec{k}) = \frac{2}{3}[4\cos(\frac{\sqrt{3}}{2}k_x)\cos(\frac{1}{2}k_y) + 2\cos(k_y) - 6]$$
$$= -k^2 + \frac{1}{16}k^4 + O(k^6) \qquad (19)$$

Details of the DFT of $L_{FV,hex}$ discretization are given in Appendix B. Both $L_{2,1}$ and $L_{FV,hex}$ discretizations may be considered isotropic because their respective rotational variances first appear at the sixth order of $k$. Fig. 2 illustrates the Fourier transform of different discretization strategies of Laplacian in 2D space. As seen in Figs 2(a) and 2(c), the 7-point stencil discretization ($L_{FV,hex}$) with hexagonal grids exhibits better isotropy than the 5-point stencil ($L_{1,0}$) with square grids. Since the Taylor expansion of Laplacian's Fourier transform terminates at the second order, any higher order terms in the expansion of a discretization method are considered error terms. By analyzing the fourth and sixth order terms in their respective Taylor expansions, we found that the $L_{FV,hex}$ discretization is more accurate and more isotropic than Ji's $L_{2,1}$ method. For both $L_{FV,hex}$ and $L_{2,1}$, the leading isotropic error occurs at the fourth order, and the leading anisotropic error occurs at the sixth order; and for both isotropic and anisotropic errors, the respective terms for $L_{FV,hex}$ discretization are smaller than those of $L_{2,1}$ (see the DFT in Appendix B).

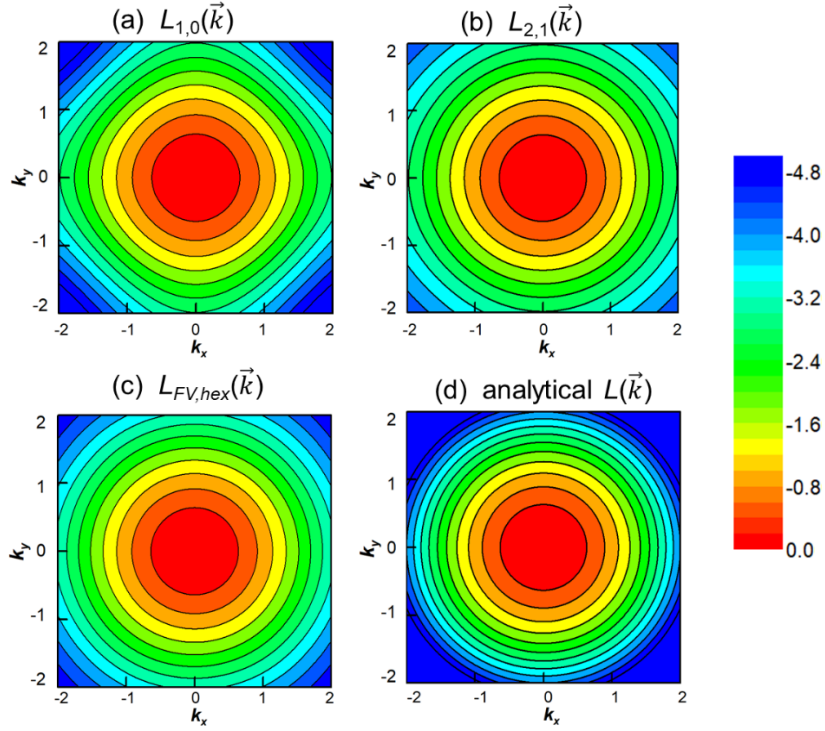

Figure 2. Fourier space contours of the 2D Laplacian with different discretization methods.

The Fourier transform of generalized divergence $\nabla \cdot (\alpha \nabla \beta)$ in continuous limit is $\tilde{D}(\vec{k}) = -2k^2$. Based on Eqn. (15), the DFT of $\nabla \cdot (\alpha \nabla \beta)$ with standard finite volume discretization in hexagonal mesh is expressed as

$$\tilde{D}_{FV,hex}(\vec{k}) = \frac{1}{3}[4\cos(\sqrt{3}k_x)\cos(k_y) + 2\cos(2k_y) - 6] \\ = -2k^2 + \frac{1}{2}k^4 + O(k^6) \qquad (20)$$

According to Eqn. (13), the DFT of the commonly used 5-point stencil discretization ($\tilde{D}_{FV,sq}$) in square grids is given by

$$\tilde{D}_{FV,sq}(\vec{k}) = \frac{1}{2}[2\cos(2k_x) + 2\cos(2k_y) - 4]$$
$$= -2k^2 + \frac{2}{3}(k_x^4 + k_y^4) + O(k^6)$$
(21)

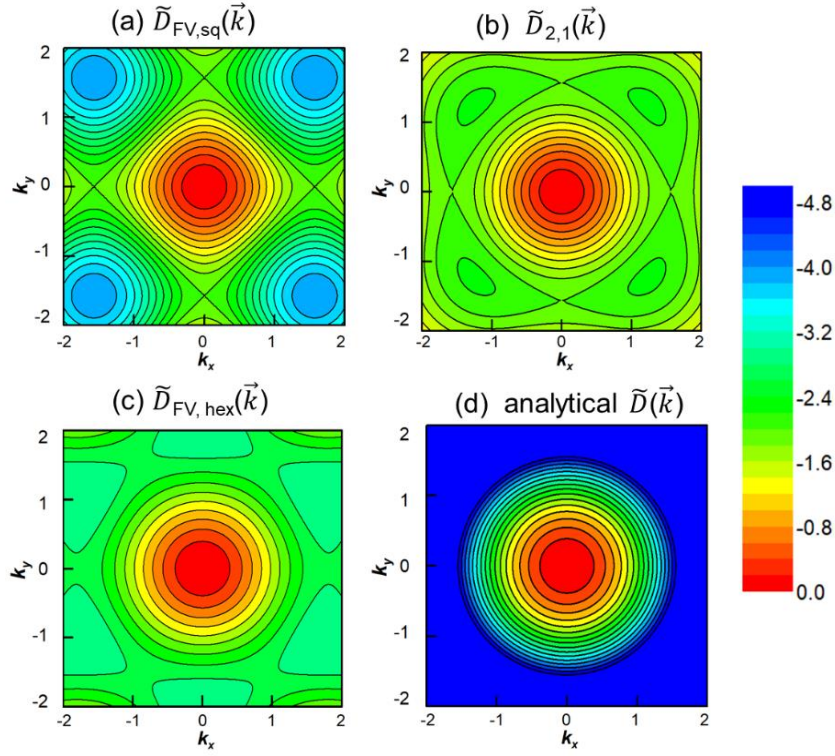

Figure 3. Fourier space contours of the 2D generalized divergence $\nabla \cdot (\alpha \nabla \beta)$ with different discretization methods.

The 5-point stencil discretization ($\tilde{D}_{FV,sq}$) in 2D square grids is anisotropic, as the rotational variance exists at the fourth order of *k*. On the other hand, the finite volume discretization in regular hexagons $\tilde{D}_{FV,hex}$ yields a scheme with better isotropy as the leading rotational variance exists at the sixth order of *k*. The DFT of different discretization methods of $\nabla \cdot (\alpha \nabla \beta)$ in Fourier space are visualized in Fig. 3. As seen in Fig. 3, both $\tilde{D}_{2,1}$ and $\tilde{D}_{FV,hex}$ exhibit better isotropy than the widely used $\tilde{D}_{FV,sq}$ discretization. The Taylor expansion to sixth order suggests that $\tilde{D}_{FV,hex}$ exhibits better accuracy and isotropy than Ji's $\tilde{D}_{2,1}$ discretization (see the DFT in Appendix B). In addition, $\tilde{D}_{FV,hex}$ is easier to be implemented with a semi-implicit algorithm, since it is a standard form of finite volume discretization and uses fewer stencil points than $\tilde{D}_{2,1}$.

Table 1. Four different schemes used for 2D PF simulations of polycrystalline solidification.

| Scheme | 2D mesh | Leading terms | Discretization | No. of stencils |
|---|---|---|---|---|
| (1) | Square | $T_1$ | $\tilde{D}_{FV,sq}$ | 5 |
|  |  | $T_2$ | $\tilde{D}_{FV,sq}$ | 5 |
|  |  | $T_3$ | Special [9] | 9 |
| (2) | Square | $T_1$ | $\tilde{D}_{FV,sq}$ | 5 |
|  |  | $T_2$ | $\tilde{D}_{FV,sq}$ | 5 |
|  |  | $\overline{T_3}$ | $\tilde{D}_{FV,sq}$ | 5 |
| (3) | Square | $T_1$ | $\tilde{D}_{2,1}$ | 9 |
|  |  | $T_2$ | $\tilde{D}_{2,1}$ | 9 |
|  |  | $\overline{T_3}$ | $\tilde{D}_{2,1}$ | 9 |
| (4) | Hexagon | $T_1$ | $\tilde{D}_{FV,hex}$ | 7 |
|  |  | $T_2$ | $\tilde{D}_{FV,hex}$ | 7 |
|  |  | $\overline{T_3}$ | $\tilde{D}_{FV,hex}$ | 7 |

## 4.2. 2D phase field simulations

Details of the semi-implicit algorithm are given in Appendix C. An open-source finite volume software called OpenFOAM was used to develop the semi-implicit PF model. The implementation of crystal axes rotation could be found elsewhere [12, 22]. Both isothermal and non-isothermal solidification of a single dendrite was simulated by using the PF model with different discretization schemes. As given in Table 1, the comparison was made by running simulations with four different schemes:

(1) the special 9-point stencil discretization [10] for the computation of the original antitrapping term in Eqn. (7), and 5-point stencil discretization $\tilde{D}_{FV,sq}$ for the computation of other leading terms in 2D square grids,

(2) the 5-point stencil discretization $\tilde{D}_{FV,sq}$ for the computation of all the leading differential terms including Eqn. (10) as approximated antitrapping current in 2D square grids,

(3) the 9-point stencil discretization $\tilde{D}_{2,1}$ for the computation of all the leading terms in 2D square grids,

(4) the 7-point stencil discretization $\tilde{D}_{FV,hex}$ for the computation of all the leading terms in hexagonal mesh.

Scheme (1) has been extensively used in previous PF simulations of alloy solidification. Scheme (2) with Eqn. (10) as approximated antitrapping term can be numerically solved with the simplest discretization. In contrast to the anisotropic schemes (1) and (2), both scheme (3) (similar to Ji *et al.* [1]) and scheme (4) (proposed in this study) employ isotropic discretization strategies. Simulations with schemes (1), (2) and (4) were performed with semi-implicit algorithm, whereas simulations with scheme (3) require an explicit implementation in our study. For schemes (1), (2), (3) with square mesh, the standard finite difference (central differencing) was applied to calculate the first order spatial derivatives in Eqn. (1), including the anisotropy strength (Eqn. (3)), the second and third terms on the right-hand side (RHS) of Eqn. (1). For scheme (4) with hexagonal mesh, the discretization of the first order spatial derivatives is given in Appendix C. Note that the discretization of first order derivatives in hexagonal mesh is a standard form for finite volume methods, which is equivalent to the isotropic discretization in finite difference lattice with regular triangles (see Eqn. (4.1) in Ref. [23]). Therefore, the simplicity of scheme (4) permits the implementation of a semi-implicit algorithm, which improves computational efficiency compared to scheme (3). Furthermore, we found that standard finite volume discretization with equilateral triangle mesh (or equivalent finite difference lattice with hexagons) can greatly reduce lattice-induced anisotropy of PF simulations compared to $\tilde{D}_{FV,sq}$ discretization. Details of the 4-point stencil discretization in triangular mesh ($\tilde{D}_{FV,tri}$) is not discussed since it is an anisotropic scheme, but one can easily derive the corresponding discretization equations on an equilateral triangle mesh.

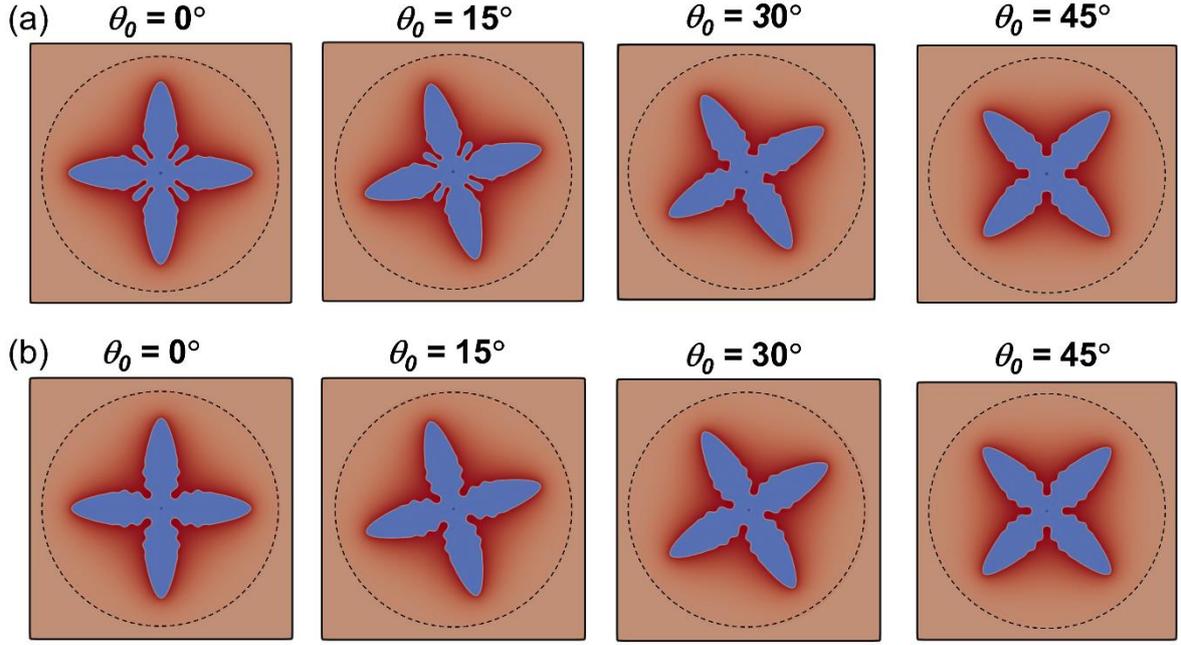

Figure 4. Schematics showing the simulated dendrite morphologies with different rotation angles $\theta_0$ by using isotropic (a) scheme (3) and (b) scheme (4), respectively. Chemical diffusivity was set to zero for regions outside the dashed circle, which is equivalent to imposing a no-flux boundary condition on the circle.

4.2.1 Isothermal solidification of single dendrite

With a partition coefficient $k_e$=0.5 and interfacial anisotropy $\varepsilon_4$=0.014, crystal growth of a single dendrite at an initial supersaturation $U = -0.4$ was simulated by using the PF model. For the testing of different schemes, a 2D square domain was discretized into square and hexagonal mesh with the same grid spacing ($h_{sq} = h_{hex}$), respectively. The 2D square domain has a size of 8800 $d_0 \times$8800 $d_0$. The liquid chemical diffusivity out of a circular region with a radius of 4000 $d_0$ was set zero, which is equivalent to imposing a no-flux Neumann boundary condition for a circular domain. At time $t = 0$, a seed with a constant radius ($r = 40\ d_0$) was placed at the center of the 2D domains. The total time of the simulations is $t_{end} = 1.28 \times 10^6$ ($d_0^2/D_l$). To examine the influences of lattice induced anisotropy, single dendrite simulations were achieved using a series of rotation angles $\theta_0$, where $\theta_0$ represents the angle between $x$-axis and preferred crystal growth direction. As seen in Fig. 4, single dendrites with different rotation angles can be used for the evaluation of different schemes. Results in Figs. 4(a) and 4(b) were computed with isotropic schemes (3) and (4), respectively. The other parameters in Fig. 4 are $W/d_0$=10 and $h/W$=0.8. For cubic alloy with four-fold symmetry in 2D, the maximum rotation angle was set as 45°.

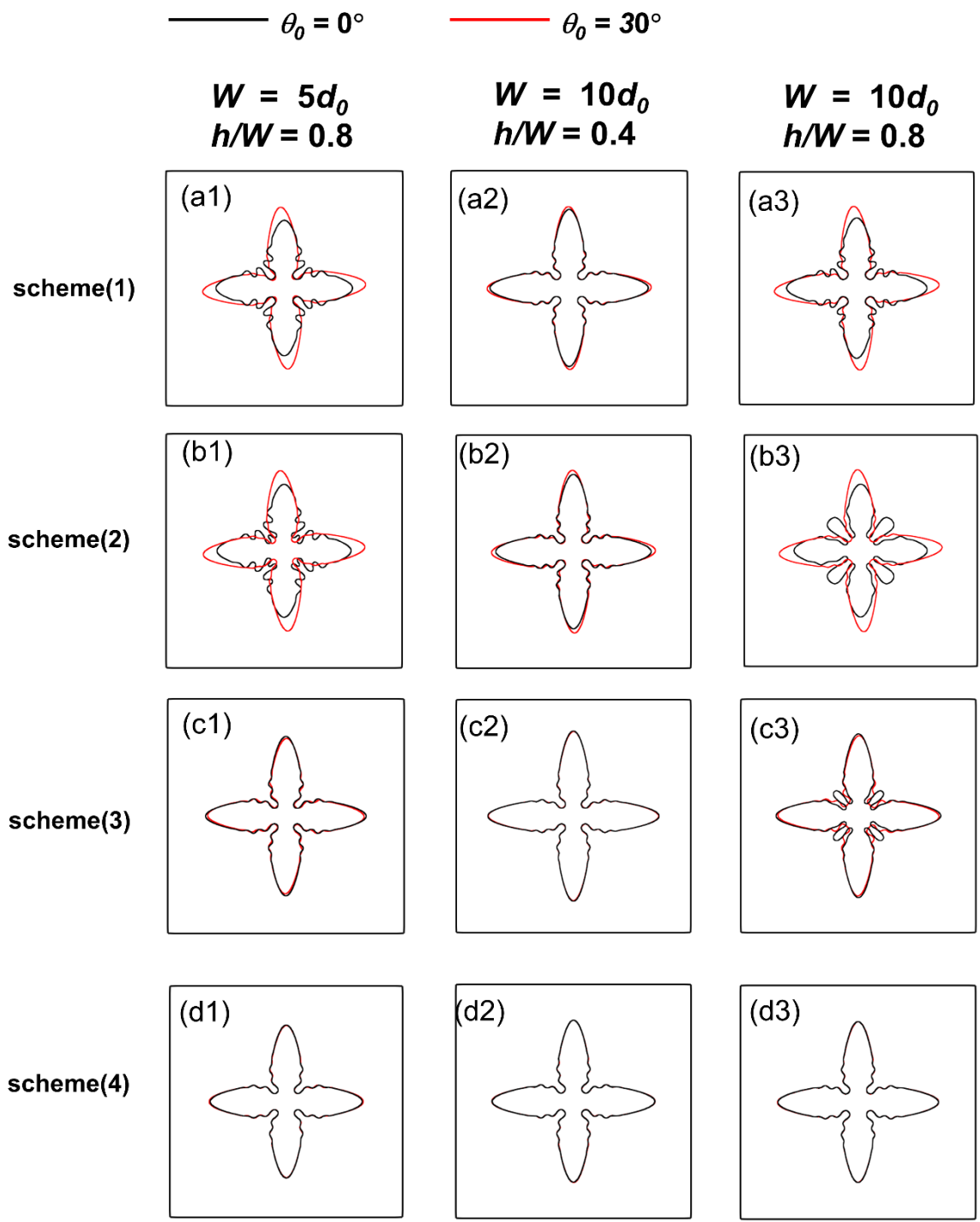

Figure 5. Comparison of the dendrite configurations with $\theta_0=0°$ and $30°$ by using different schemes. The single dendrites with $\theta_0=30°$ were rotated by $-30°$ to examine the rotational invariance as compared to $\theta_0=0°$.

Solidification of single dendrite with different rotation angles $\theta_0=0°$ and 30° was compared in Fig. 5. Results with $\theta_0=30°$ were rotated with an angle $-\theta_0$ to examine the rotational dependences of different schemes. For an ideal isotropic scheme, the dendrite configuration should be rotationally invariant with $\theta_0=0°$ and 30°. As seen in Figs. 5(a1)-(a3) and 5(b1)-(b3), the anisotropic schemes (1) and (2) clearly introduce rotational variance for polycrystalline simulations, especially for those cases where the grid size is close to the interfacial thickness ($h/W=0.8$). Compared to schemes (1) and (2), results using scheme (3) with 9-point stencil discretization $\tilde{D}_{2,1}$ show less rotational variance. Only small discrepancies between $\theta_0=0°$ and 30° was found for results using scheme (3). Simulations using hexagonal mesh and scheme (4) exhibit the best performance in terms of rotational invariance. With the application of $\tilde{D}_{FV,hex}$ discretization in scheme (4), the dendrite configurations are almost indistinguishable between $\theta_0=0°$ and 30°.

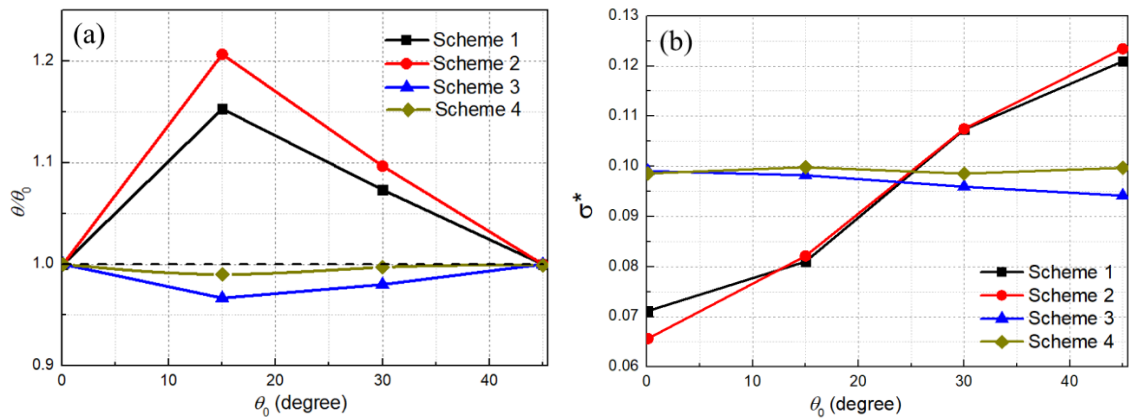

Figure 6. Comparison of (a) scaled growth direction and (b) tip selection parameters for PF simulations of isothermal solidification with different schemes. The solid lines are guides for the eye.

Following Ji et al. [1], the performance of different schemes was evaluated by examining another two criteria, including the scaled growth direction $\theta/\theta_0$ and the scaling parameter $\sigma*$. The angle $\theta$ represents the actual growth direction of dendrite solidification, where $\theta_0$ is the growth direction imposed by the anisotropy function in the PF model. Therefore, $\theta/\theta_0$ =1 would be achieved for an ideal isotropic scheme. In this work, $\theta$ was determined by rotating the dendrite configurations by ($-\theta$) such that the crystal axes coincide with the $x$ and $y$ axes. According to the linear solvability theory [24], the scaling parameter $\sigma*=2D_l d_0/\rho^2 V$ (where $V$ is tip growth velocity and $\rho$ is tip radius) should reach a constant with time at low undercooling. In our simulations, we measured the tip radius and tip growth velocity of different cases at $t = 1.28\times10^6$ ($d_0^2/D_l$). All the PF simulations shown in Fig. 6 were conducted with $W/d_0=10$ and $h/W=0.8$. As shown in Fig. 6(a), the accuracy of scheme (4) is better than the isotropic scheme (3) in terms of the scaled growth direction, for which the maximum errors are 4.0% and 1.0%, respectively, for scheme (3) and scheme (4). In Fig. 6(b), both schemes (3) and (4) exhibit relatively small variances in the scaling parameter, $\sigma*$, compared to the anisotropic schemes (1) and (2). The standard deviations of $\sigma*$ for scheme (3) and (4) are, respectively, $2.2\times10^{-3}$ and $6.9\times10^{-4}$, indicating that scheme (4) is more isotropic than scheme (3). Therefore, the proposed isotropic discretization $\tilde{D}_{FV,hex}$ exhibits better rotational invariance and better accuracy than other discretization methods for 2D simulations of polycrystalline solidification.

Due to the simple form of $\tilde{D}_{FV,hex}$ discretization, one can easily implement it with a semi-implicit algorithm to improve the computational efficiency. The computational performance of explicit approach (scheme (3)) and semi-implicit approach (scheme (4)) are compared and summarized in Table 2. As given in Table 2, the semi-implicit algorithm with $\tilde{D}_{FV,hex}$ discretization (scheme (4)) is considerably more efficient than the explicit approach with $\tilde{D}_{2,1}$ discretization (scheme (3)). For the semi-implicit approach, the time step constraint is increased by one to two orders of magnitudes. Simulations with semi-implicit algorithms and anisotropic schemes (1) or (2) are a bit faster than the cases with scheme (4) (not shown in the table). While $\tilde{D}_{FV,hex}$ is slightly less efficient than the anisotropic discretization $\tilde{D}_{FV,sq}$, it is more accurate than other discretization methods and can be readily implemented with semi-implicit algorithms.

Table 2. Computational performances of schemes (3) and (4) with explicit and semi-implicit algorithms for PF simulations of isothermal solidification. The PF simulations were conducted on a high performance cluster with 48 Intel® Xeon Phi 7210 processors.

| Parameters | Scheme | Algorithm | timestep ($h^2/D_l$) | Run time (hours) |
|---|---|---|---|---|
| $W/d_0 = 10, h/W = 0.8$ | 3 | Explicit | 0.2 | 5.3 |
| | 4 | Semi-implicit | 3.125 | 0.58 |
| $W/d_0 = 10, h/W = 0.4$ | 3 | Explicit | 0.2 | 68 |
| | 4 | Semi-implicit | 12.5 | 2.7 |
| $W/d_0 = 5, h/W = 0.8$ | 3 | Explicit | 0.2 | 68 |
| | 4 | Semi-implicit | 3.125 | 9.0 |

4.2.1 Directional solidification

Additionally, the performances of different schemes were examined by simulating crystal growth under a fixed temperature gradient $G$ and a pulling velocity $V_p$. With different crystal growth angle $\theta_0$, directional solidification of a single crystal was achieved by using a relatively large $V_p$. Following Ji et al. [1], we compared the characteristic length scale $\delta$ obtained from PF simulations and the analytical solutions of the Ivantsov-solvability theory [25]. The characteristic length scale $\delta$ demonstrates the distance between the dendrite tips of well-orientated grain ($\theta_0=0°$) and misoriented grains ($\theta_0>0°$). In our PF simulations, the characteristic length scale was defined as [1, 25]

$$\delta = (\Delta_1 - \Delta_0)l_T \tag{22}$$

where $\Delta_1$ and $\Delta_0$ respectively represent dimensionless tip undercooling $\Delta = (T_l - T_{tip})/\Delta T_0$ of misoriented and well-oriented grains, $T_l$ is the liquidus temperature, $\Delta T_0$ is the freezing range, $T_{tip}$ is tip temperature, $l_T=\Delta T_0/G$ is the thermal length. For directional solidification simulations, the dimensionless tip undercooling was measured when the tip growth velocity reached a steady-state value of $V_p$.

Except for the time step used, the simulation parameters in our study are identical to those used in Ji et al. [1, 25]. For scheme (3) with explicit algorithm, the time step was set as $\Delta t=1\times10^{-5}$ s. For schemes (1), (2) and (4) with semi-implicit algorithm, the time step was set as $\Delta t=8\times10^{-5}$ s. Simulation results with scheme (4) showing the directional solidification of a model system with material properties characteristic of SCN-1.3 wt.% acetone alloy are illustrated in Fig. 7(a). To ensure that the dendrite tip is not affected by its neighbours due to periodic boundary conditions along the $x$ direction, our PF simulations used a domain size of $L_x \times L_y = 384 \times 480$ μm$^2$. Therefore, the smallest primary dendrite spacing is 271.5 μm ($\theta_0=45°$), which is much larger than the diffusion length $l_D=15$ μm [1].

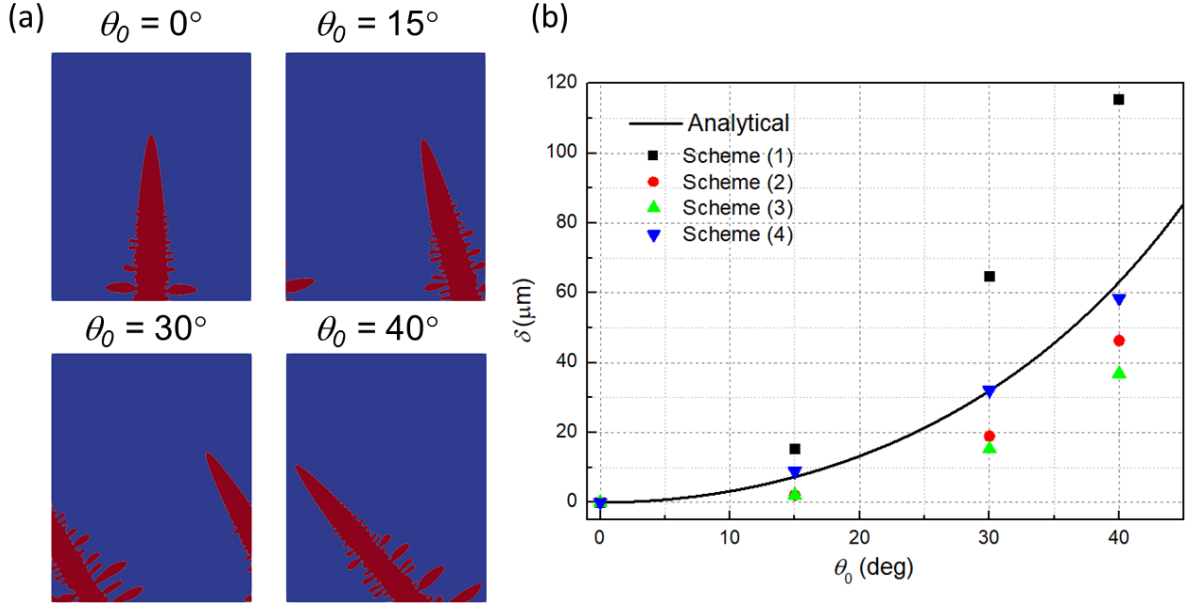

Figure 7. (a) 2D simulations of directional solidification of single dendrite with scheme (4) and different growth angle $\theta_0$, and (b) the characteristic length as a function of $\theta_0$ with different schemes.

As shown in Fig. 7(b), our proposed isotropic scheme (4) shows the best agreement with Ivantsov's prediction among all the schemes compared in this work. With the implementation of semi-implicit algorithm, the computational efficiency of scheme (4) is also better than scheme (3), as compared in Table 3. To quantitatively simulate growth competition of well-oriented and misoriented grains in 2D, it is thus recommended to use the proposed discretization $\tilde{D}_{FV,hex}$ with hexagonal mesh and semi-implicit algorithm.

Table 3. Computational performances of schemes (3) and (4) with explicit and semi-implicit algorithms for PF simulations of directional solidification. The PF simulations were conducted on a high performance cluster with 48 Intel® Xeon Phi 7210 processors.

| Scheme | Algorithm | timestep (s) | Run time (hours) |
|--------|-----------|--------------|------------------|
| 3 | Explicit | $1\times10^{-5}$ | 110.6 |
| 4 | Semi-implicit | $8\times10^{-5}$ | 29.8 |

**Conclusions**

In this study, we proposed a 2D isotropic method utilizing hexagonal mesh to discretize the leading differential terms in the governing equations of the PF model of alloy solidification. The rotational dependences of different discretization methods, including $\tilde{D}_{FV,sq}$, $\tilde{D}_{2,1}$ and $\tilde{D}_{FV,hex}$, were analysed by the known method of discrete Fourier transformation. The DFT analysis suggests a better level of accuracy of the proposed discretization $\tilde{D}_{FV,hex}$ compared to the conventional $\tilde{D}_{FV,sq}$ and the $\tilde{D}_{2,1}$ method by Ji *et al.* [1]. The proposed isotropic discretization can be implemented in a finite volume mesh with hexagonal cells, which is different from the application of $\tilde{D}_{FV,sq}$ or $\tilde{D}_{2,1}$ discretization on a square mesh. Since the proposed method is a standard form of finite volume discretization, it is straightforward to apply the discretization $\tilde{D}_{FV,hex}$ together with a semi-implicit algorithm to solve the PF equations.

For 2D alloy solidification, both isothermal and non-isothermal solidification of a single dendrite with various rotation angles were simulated by using different schemes. Results suggest that the proposed discretization can drastically reduce lattice-induced anisotropy in PF simulations. The accuracy of the proposed method in this work performs better than the $\tilde{D}_{2,1}$ discretization proposed by Ji *et al*. [1]. Moreover, the semi-implicit algorithm with standard form of finite volume discretization can greatly reduce the time step constraints and improves the computational efficiency. Therefore, the proposed isotropic discretization can be potentially useful for accurate prediction of 2D PF simulations, such as dendrite sedimentation, directional solidification, rapid solidification during additive manufacturing.

**Declaration of competing interest**

The authors declare that there is no conflict of interest.

**Acknowledgment**

The authors would like to thank Dr. Kaihua Ji for the discussion of Fourier space analysis of discretization schemes.

SSQ acknowledges financial support from A*STAR, Singapore under the Structural Metals and Alloys Program (Grant No.:A18B1b0061).

This work was also supported by the A*STAR Computational Resource Centre through the use of its high performance computing facilities.

**Appendix A.** $L_{2,1}$ **and** $\tilde{D}_{2,1}$ **schemes in 2D square mesh [1]**

For the Laplacian in 2D square mesh, a 5-point stencil is commonly used with contribution from a point and its first nearest neighbors. It is referred as $L_{FV,sq}$ (identical to $L_{1,0}$ in Ref.[1]) in this work, and its form is given as

$$L_{1,0} = \frac{1}{h_{sq}^2}\left[(\beta_{(1,0)} + \beta_{(0,1)} + \beta_{(-1,0)} + \beta_{(0,-1)} - 4\beta_{(0,0)})\right] \quad (A1)$$

Another base discretization of Laplacian using the second nearest neighbours is denoted as $L_{0,1}$, and is expressed as

$$L_{0,1} = \frac{1}{2h_{sq}^2}\left[(\beta_{(1,1)} + \beta_{(-1,1)} + \beta_{(-1,-1)} + \beta_{(1,-1)} - 4\beta_{(0,0)})\right] \tag{A2}$$

The isotropic discretization of Laplacian in 2D square mesh is denoted as $L_{2,1}$, which combines the two base discretization $L_{1,0}$ and $L_{0,1}$. Accordingly, the discretization form of $L_{2,1}$ is

$$L_{2,1} = \frac{2}{3}L_{1,0} + \frac{1}{3}L_{0,1} \tag{A3}$$

Similarly, the generalized divergence $\nabla \cdot (\alpha \nabla \beta)$ in 2D square mesh involves two base discretization components $\tilde{D}_{1,0}$ and $\tilde{D}_{0,1}$. The base discretization $\tilde{D}_{1,0}$ is expressed as

$$\tilde{D}_{1,0} = \frac{1}{h_{sq}}\left[F_{(\frac{1}{2},0)} + F_{(0,\frac{1}{2})} + F_{(-\frac{1}{2},0)} + F_{(0,-\frac{1}{2})}\right] \tag{A4}$$

Where $F$ represents the flux and is given by

$$\begin{aligned}
F_{(\frac{1}{2},0)} &= \frac{1}{h_{sq}}\left[\alpha_{(1,0)} + \alpha_{(0,0)} + \alpha_{(\frac{1}{2},\frac{1}{2})} + \alpha_{(\frac{1}{2},-\frac{1}{2})}\right]\left[\beta_{(1,0)} - \beta_{(0,0)}\right] \\
F_{(0,\frac{1}{2})} &= \frac{1}{h_{sq}}\left[\alpha_{(0,1)} + \alpha_{(0,0)} + \alpha_{(\frac{1}{2},\frac{1}{2})} + \alpha_{(-\frac{1}{2},\frac{1}{2})}\right]\left[\beta_{(0,1)} - \beta_{(0,0)}\right] \\
F_{(-\frac{1}{2},0)} &= \frac{1}{h_{sq}}\left[\alpha_{(0,0)} + \alpha_{(-1,0)} + \alpha_{(-\frac{1}{2},\frac{1}{2})} + \alpha_{(-\frac{1}{2},-\frac{1}{2})}\right]\left[\beta_{(-1,0)} - \beta_{(0,0)}\right] \\
F_{(0,-\frac{1}{2})} &= \frac{1}{h_{sq}}\left[\alpha_{(0,0)} + \alpha_{(0,-1)} + \alpha_{(\frac{1}{2},-\frac{1}{2})} + \alpha_{(-\frac{1}{2},-\frac{1}{2})}\right]\left[\beta_{(0,-1)} - \beta_{(0,0)}\right]
\end{aligned} \tag{A5}$$

The $\alpha$ at the off-lattice position is computed as

$$\begin{aligned}
\alpha_{(\frac{1}{2},\frac{1}{2})} &= \frac{1}{4h_{sq}}\left[\alpha_{(1,0)} + \alpha_{(1,1)} + \alpha_{(0,1)} + \alpha_{(0,0)}\right] \\
\alpha_{(\frac{1}{2},-\frac{1}{2})} &= \frac{1}{4h_{sq}}\left[\alpha_{(1,0)} + \alpha_{(1,-1)} + \alpha_{(0,-1)} + \alpha_{(0,0)}\right] \\
\alpha_{(-\frac{1}{2},-\frac{1}{2})} &= \frac{1}{4h_{sq}}\left[\alpha_{(-1,0)} + \alpha_{(-1,-1)} + \alpha_{(0,-1)} + \alpha_{(0,0)}\right] \\
\alpha_{(-\frac{1}{2},\frac{1}{2})} &= \frac{1}{4h_{sq}}\left[\alpha_{(-1,0)} + \alpha_{(-1,1)} + \alpha_{(0,1)} + \alpha_{(0,0)}\right]
\end{aligned} \tag{A6}$$

The base discretization $\tilde{D}_{0,1}$ is expressed as

$$\tilde{D}_{0,1} = \frac{1}{\sqrt{2}h_{sq}}\left[F_{(\frac{1}{2},\frac{1}{2})} + F_{(-\frac{1}{2},\frac{1}{2})} + F_{(\frac{1}{2},-\frac{1}{2})} + F_{(-\frac{1}{2},-\frac{1}{2})}\right] \quad (A7)$$

Where the flux is given by

$$F_{(\frac{1}{2},\frac{1}{2})} = \frac{1}{4\sqrt{2}h_{sq}}\left[\alpha_{(1,0)} + \alpha_{(0,0)} + \alpha_{(0,1)} + \alpha_{(1,0)}\right]\left[\beta_{(1,1)} - \beta_{(0,0)}\right]$$

$$F_{(-\frac{1}{2},\frac{1}{2})} = \frac{1}{4\sqrt{2}h_{sq}}\left[\alpha_{(-1,1)} + \alpha_{(0,0)} + \alpha_{(0,1)} + \alpha_{(-1,0)}\right]\left[\beta_{(-1,1)} - \beta_{(0,0)}\right]$$

$$F_{(\frac{1}{2},-\frac{1}{2})} = \frac{1}{4\sqrt{2}h_{sq}}\left[\alpha_{(1,-1)} + \alpha_{(0,0)} + \alpha_{(1,0)} + \alpha_{(0,-1)}\right]\left[\beta_{(1,-1)} - \beta_{(0,0)}\right]$$

$$F_{(-\frac{1}{2},-\frac{1}{2})} = \frac{1}{4\sqrt{2}h_{sq}}\left[\alpha_{(-1,-1)} + \alpha_{(0,0)} + \alpha_{(-1,0)} + \alpha_{(0,-1)}\right]\left[\beta_{(-1,-1)} - \beta_{(0,0)}\right]$$

(A8)

The linear combination of $\tilde{D}_{1,0}$ and $\tilde{D}_{0,1}$ gives the isotropic form of $\tilde{D}_{2,1}$ in square mesh, and is expressed by

$$\tilde{D}_{2,1} = \frac{2}{3}\tilde{D}_{1,0} + \frac{1}{3}\tilde{D}_{0,1} \quad (A9)$$

**Appendix B. Discrete Fourier transform of isotropic discretization schemes**

Using the properties of Fourier transform, the Fourier transform of Laplacian $\nabla^2 \beta$ in continuum limit is given as

$$\frac{F[\nabla^2 \beta(r)]}{F[\beta(r)]} = \frac{-k^2 F[\beta(r)]}{F[\beta(r)]} = -k^2 \quad (B1)$$

For discrete Fourier transform, we can use the properties:

$$F[f(x+x_0, y+y_0)] = e^{-ik_x x_0} e^{-ik_y y_0} F[f(x,y)] \quad (B2)$$

$$F[f(x)g(x)] = F[f(x)]F[g(x)] \quad (B3)$$

According to Eqn. (17), the DFT of $L_{FV,hex}$ is given by

$$\frac{F[L_{FV,hex}]}{F[\beta(r)]} = \frac{2F[\beta_{(0,0)}]}{3F[\beta(r)]}\left\{e^{-i\frac{\sqrt{3}}{2}k_x}e^{-i\frac{1}{2}k_y} + e^{-ik_y} + e^{i\frac{\sqrt{3}}{2}k_x}e^{-i\frac{1}{2}k_y} + e^{i\frac{\sqrt{3}}{2}k_x}e^{i\frac{1}{2}k_y} + e^{ik_y} + e^{-i\frac{\sqrt{3}}{2}k_x}e^{i\frac{1}{2}k_y} - 6\right\}$$

$$= \frac{2}{3}[4\cos(\frac{\sqrt{3}}{2}k_x)\cos(\frac{1}{2}k_y) + 2\cos(k_y) - 6] \tag{B4}$$

$$= -k^2 + \frac{1}{16}k^4 + \frac{k^6}{5760}[-10 + \cos(6\theta_k)] + O(k^8)$$

where $\cos\theta_k = \dfrac{k_x}{\sqrt{k_x^2 + k_y^2}}$.

The DFT of $L_{2,1}$ is

$$\frac{F[L_{2,1}]}{F[\beta(r)]} = \frac{2}{3}[2\cos(k_x) + 2\cos(k_y) + \cos(k_x)\cos(k_y) - 5]$$

$$= -k^2 + \frac{1}{12}k^4 + \frac{k^6}{1440}[-5 + \cos(4\theta_k)] + O(k^8) \tag{B5}$$

The Fourier transform of generalized divergence $\nabla\cdot(\alpha\nabla\beta)$ in continuum limit is given as

$$\frac{F[\nabla\cdot(\alpha(r)\nabla\beta(r))]}{F[\alpha(r)\beta(r)]} = \frac{F[\nabla\alpha(r)\cdot\nabla\beta(r) + \alpha(r)\nabla^2\beta(r)]}{F[\alpha(r)]F[\beta(r)]} = -2k^2 \tag{B6}$$

According to Eqn. (15), the DFT of $\tilde{D}_{FV,hex}$ is

$$\frac{F[\tilde{D}_{FV,hex}]}{F[\alpha(r)\beta(r)]} = \frac{F[\alpha_{(0,0)}]F[\beta_{(0,0)}]}{3F[\alpha(r)]F[\beta(r))]} *$$

$$\left\{ (e^{-i\frac{\sqrt{3}}{2}k_x}e^{-i\frac{1}{2}k_y}+1)(e^{-i\frac{\sqrt{3}}{2}k_x}e^{-i\frac{1}{2}k_y}-1)+(e^{-ik_y}+1)(e^{-ik_y}-1) \right.$$

$$+(e^{i\frac{\sqrt{3}}{2}k_x}e^{-i\frac{1}{2}k_y}+1)(e^{i\frac{\sqrt{3}}{2}k_x}e^{-i\frac{1}{2}k_y}-1)+(e^{i\frac{\sqrt{3}}{2}k_x}e^{i\frac{1}{2}k_y}+1)(e^{i\frac{\sqrt{3}}{2}k_x}e^{i\frac{1}{2}k_y}-1)$$

$$\left. +(e^{ik_y}+1)(e^{ik_y}-1)+(e^{-i\frac{\sqrt{3}}{2}k_x}e^{i\frac{1}{2}k_y}+1)(e^{-i\frac{\sqrt{3}}{2}k_x}e^{i\frac{1}{2}k_y}-1) \right\}$$

$$= \frac{1}{3}\left\{ (e^{-i\sqrt{3}k_x}e^{-ik_y}-1)+(e^{-i2k_y}-1)+(e^{i\sqrt{3}k_x}e^{-ik_y}-1) \right.$$

$$\left. +(e^{i\sqrt{3}k_x}e^{ik_y}-1)+(e^{i2k_y}-1)+(e^{-i\sqrt{3}k_x}e^{ik_y}-1) \right\}$$

$$= \frac{1}{3}[4\cos(\sqrt{3}k_x)\cos(k_y)+2\cos(2k_y)-6]$$

$$= -2k^2 + \frac{1}{2}k^4 + \frac{k^6}{180}[-10+\cos(6\theta_k)]+O(k^8) \tag{B7}$$

Based on Ji's study[1], the DFT of $\tilde{D}_{2,1}$ is

$$\frac{F[\tilde{D}_{2,1}]}{F[\alpha(r)\beta(r)]} = \frac{2}{3}[\frac{1}{8}\cos(2k_x-k_y)+\frac{1}{8}\cos(2k_x-k_y)+\frac{1}{8}\cos(2k_x-k_y)+\frac{1}{8}\cos(2k_x-k_y)$$

$$+\frac{3}{4}\cos(2k_x)+\frac{3}{4}\cos(2k_y)-\frac{1}{4}\cos(k_x)-\frac{1}{4}\cos(k_y)-\frac{3}{2}]$$

$$+\frac{1}{3}[\frac{1}{2}\cos(2k_x)\cos(2k_y)+\frac{1}{2}\cos(k_x)\cos(2k_y)+\frac{1}{2}\cos(2k_x)\cos(k_y) \tag{B8}$$

$$-\frac{1}{2}\cos(k_x)-\frac{1}{2}\cos(k_y)-\frac{1}{2}]$$

$$= -2k^2 + \frac{2}{3}k^4 + \frac{k^6}{1440}[-145+17\cos(4\theta_k)]+O(k^8)$$

**Appendix C. Semi-implicit algorithm with isotropic discretization scheme (4) in hexagonal mesh**

We developed semi-implicit algorithms to solve the governing equations of the PF model. Discretization schemes of the terms in Eqns. (1), (2), and (10) are given as follows.

The left hand side (LHS) of Eqn. (1) is

$$\tau_0 a_s^2[1+(1-k_e)U]\frac{\partial \phi}{\partial t} = \tau_0 (a_s^t)^2[1+(1-k_e)U^t]\frac{\phi_p^{t+\Delta t}-\phi_p^t}{\Delta t} \tag{C1}$$

Computing anisotropy strength is based on the gradient of $\phi$. The gradient computation is given by

$$\nabla \phi = \frac{2}{3h_{hex}} \sum_{N=1}^{6} \vec{n}_{f,N} \frac{(\phi_N^t + \phi_P^t)}{2} \tag{C2}$$

where $\vec{n}_{f,N}$ is the unit normal vector of the hexagon edge. Take Fig. 1(b) for example, the gradient is

$$\begin{aligned}\nabla \phi = &\frac{1}{3h_{hex}}[(\frac{\sqrt{3}}{2}\vec{i},\frac{1}{2}\vec{j})\cdot(\phi_{(0,0)}+\phi_{(\frac{\sqrt{3}}{2},\frac{1}{2})})]\\
&+\frac{1}{3h_{hex}}[(0\vec{i},\vec{j})\cdot(\phi_{(0,0)}+\phi_{(0,1)})]\\
&+\frac{1}{3h_{hex}}[(-\frac{\sqrt{3}}{2}\vec{i},\frac{1}{2}\vec{j})\cdot(\phi_{(0,0)}+\phi_{(-\frac{\sqrt{3}}{2},\frac{1}{2})})]\\
&+\frac{1}{3h_{hex}}[(-\frac{\sqrt{3}}{2}\vec{i},-\frac{1}{2}\vec{j})\cdot(\phi_{(0,0)}+\phi_{(-\frac{\sqrt{3}}{2},-\frac{1}{2})})]\\
&+\frac{1}{3h_{hex}}[(\frac{\sqrt{3}}{2}\vec{i},-\frac{1}{2}\vec{j})\cdot(\phi_{(0,0)}+\phi_{(\frac{\sqrt{3}}{2},-\frac{1}{2})})]\\
&+\frac{1}{3h_{hex}}[(0\vec{i},-\vec{j})\cdot(\phi_{(0,0)}+\phi_{(0,-1)})]\end{aligned} \tag{C3}$$

The anisotropy strength can be computed by substituting the $x$ and $y$ component of $\nabla \phi$ into Eqn. (3).

The first term on the right hand side (RHS) of Eqn. (1) is expressed as

$$W^2 \nabla \cdot (a_s^2 \nabla \phi) = \frac{W^2}{3h_{hex}^2} \sum_{N=1}^{6}\left[(a_{s,N}^t)^2 + (a_{s,P}^t)^2\right](\phi_N^{t+\Delta t}-\phi_P^{t+\Delta t}) \tag{C4}$$

The second and third term can also be expressed as

$$W^2 \partial_x (|\nabla \phi|^2 a_s \frac{\partial a_s}{\partial \phi_x}) = W^2 \frac{\partial}{\partial x}[a_s \cdot 16\varepsilon_4 \cdot \frac{\phi_x^3(\phi_x^2 + \phi_y^2) - \phi_x(\phi_x^4 + \phi_y^4)}{(\phi_x^2 + \phi_y^2)^2}] \tag{C5}$$

$$W^2 \partial_y (|\nabla \phi|^2 a_s \frac{\partial a_s}{\partial \phi_y}) = W^2 \frac{\partial}{\partial y}[a_s \cdot 16\varepsilon_4 \cdot \frac{\phi_y^3(\phi_x^2 + \phi_y^2) - \phi_y(\phi_x^4 + \phi_y^4)}{(\phi_x^2 + \phi_y^2)^2}] \tag{C6}$$

For the discretization of Eqns. (C5) and (C6), one can use a similar way in Eqn. (C2) to compute the gradient of the terms within the square brackets. The $x$ component in (C5) and the $y$ component in (C6) are substituted into the discretized equations in OpenFOAM.

The remaining terms in Eqn (1) is given as

$$(\phi - \phi^3) - \lambda(1-\phi^2)^2 \left[ U + \frac{T-T_s}{|m_l|c_l^0(1-k_e)} \right] = \left[ \phi_P^t - (\phi_P^t)^3 \right] - \lambda \left[ 1 - (\phi_P^t)^2 \right]^2 \left[ U_P^t + \frac{T-T_s}{|m_l|c_l^0(1-k_e)} \right] \tag{C7}$$

For the solute transport equation, the LHS of Eqn. (2) is computed as

$$\frac{1}{2}[1 + k_e - (1-k_e)\phi]\frac{\partial U}{\partial t} = \frac{1}{2}[1 + k_e - (1-k_e)\phi_P^{t+\Delta t}]\frac{U_P^{t+\Delta t} - U_P^t}{\Delta t} \tag{C8}$$

The first term on the RHS is given as

$$\nabla \cdot (\frac{1-\phi}{2} D_l \nabla U) = \frac{D_l}{6h_{tri}^2} \sum_{N=1}^{6} (2 - \phi_N^{t+\Delta t} - \phi_P^{t+\Delta t})(U_N^{t+\Delta t} - U_P^{t+\Delta t}) \tag{C9}$$

The second term on the RHS is described by

$$\frac{1}{2}[1 + (1-k_e)U]\frac{\partial \phi}{\partial t} = \frac{1}{2}[1 + (1-k_e)U_P^t]\frac{\phi_P^{t+\Delta t} - \phi_P^t}{\Delta t} \tag{C10}$$

Finally, the approximated antitrapping term is explicitly computed with a similar manner in Eqn. (C4). For the approximated antitrapping term, the $\alpha$ field in generalized $\nabla \cdot (\alpha \nabla \beta)$ is

$$\alpha_P^t = \frac{\left[1+(1-k_e)U_P^t\right]W^2}{2} \cdot \frac{1}{\left|1-\left(\phi_P^{t+\Delta t}\right)^2\right|+0.001} \cdot \frac{\phi_P^{t+\Delta t}-\phi_P^t}{\Delta t} \tag{C11}$$

Accordingly, the discretization of the approximated antitrapping term is given as

$$\tilde{A}(\vec{r}) = \nabla \cdot (\alpha \nabla \phi) = \frac{1}{3h_{tri}^2}\sum_{N=1}^{6}\left[\alpha_P^t+\alpha_N^t\right](\phi_N^{t+\Delta t}-\phi_P^{t+\Delta t}) \tag{C12}$$